# The Strike for Black Lives: Growth of community in the Cal-Bridge program

The engagement inspired by the June 2020 call for action in academia has blossomed into change-promoting groups that the bridge program's scholars both participate in and lead.

**Alexander L. Rudolph**

**Carol Hood**

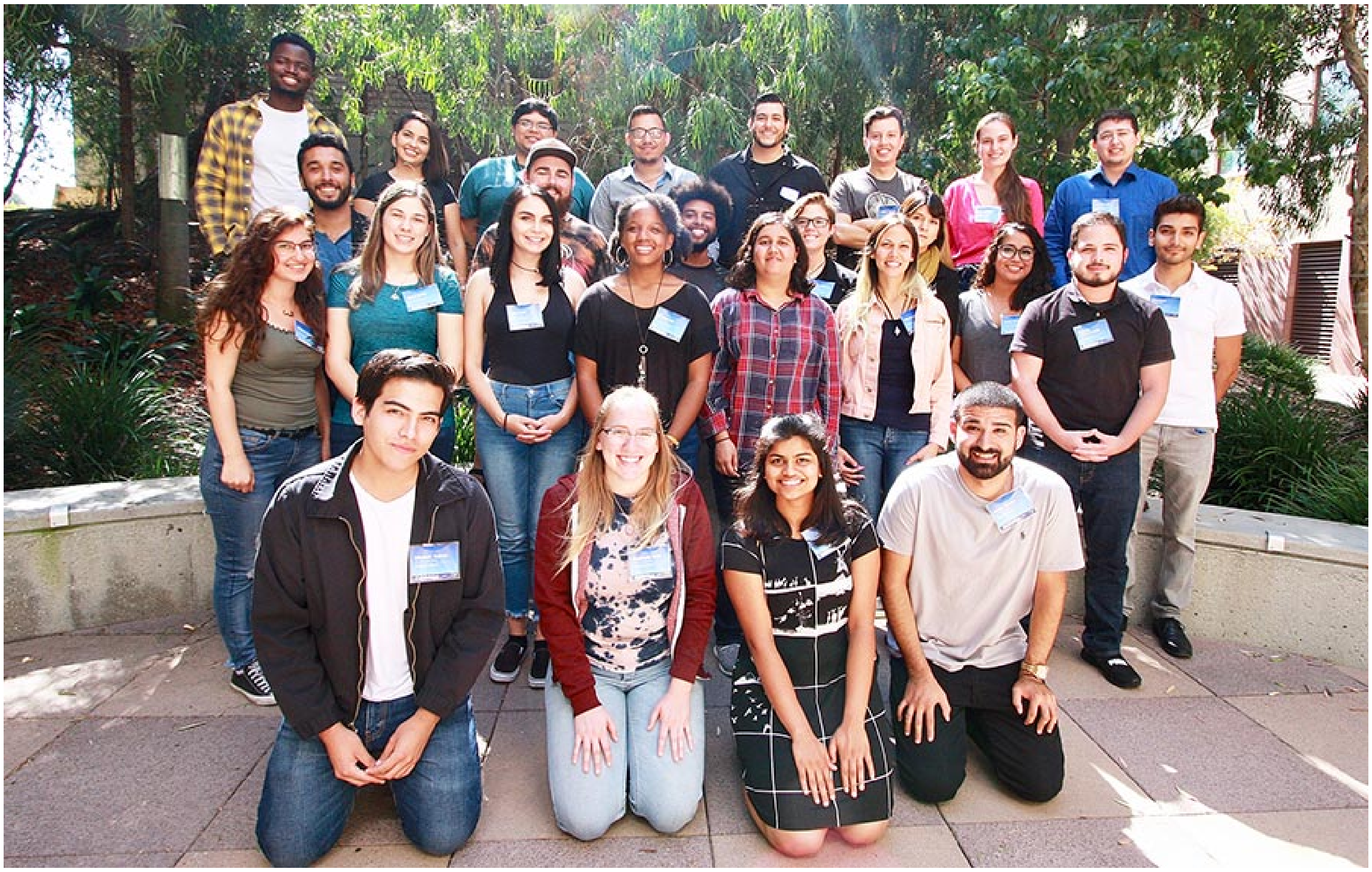

Cal-Bridge South scholars at fall 2018 orientation. Credit: Cal-Bridge program

Events of 2020, including but not limited to the killings of George Floyd, Ahmaud Arbery, and Breonna Taylor, sparked a heightened awareness across the US of the racial inequities that permeate our society. Far from being isolated, events like those represent a pattern of systemic racism that exists everywhere, including in academia. In June 2020, a group of physicists named Particles for Justice called for a general STEM strike, the Strike for Black Lives. They were joined in that call by two other groups, Shutdown STEM and Vanguard STEM. The call was both in response to the continued murder of Black people at the hands of police and vigilantes and in recognition of the systemic anti-Black racism embedded in academia.

The purpose of the strike was for individuals and organizations to reflect on what concrete actions they could take to combat anti-Black racism in academia and our society. Whereas Black academics were encouraged to “take the day to do whatever nourishes their hearts,” organizers emphasized that “the strike is not a ‘day off’ for non-Black scientists, but a day to engage in academia’s core mission to build a better society for everyone.”

We report here on the outcome of meetings associated with the strike among the community of the Cal-Bridge program, which is designed to help rising juniors in the California State University (CSU) system successfully matriculate to STEM PhD programs. There was a particularly strong sense of community engagement among Cal-Bridge alumni. Our experiences may help those in other institutions to turn their ideas for change into concrete actions.

## The importance of community

It is well known that human beings need the affiliation and attachment that come from belonging to a community; people survive and thrive when they feel socially connected.[1,2] In the academic context, Mica Estrada, Alegra Eroy-Reveles, and John Matsui have described the key role that social inclusion and a sense of belonging play in promoting persistence among historically underrepresented (HU) students in academia, including underrepresented minorities (URMs), women in STEM, LGBTQ+ students, first-generation students, and students with disabilities, among others.[3] In particular, the development of discipline identity, a form of social identity, has been shown to be a strong predictor of persistence in STEM.[4–6]

Historically underrepresented students’ sense of belonging is often obstructed by their perception that academia can be hostile or unwelcoming, with social norms that differ from their previous cultural experiences.[3,7] Many HU students, especially URMs, are raised in a more collectivist culture than is acknowledged in academia, which tends to focus on individual achievement and often underplays the key role of collective work.[8,9] The obstruction HU scholars face can take the form of racial tension, microaggressions, and insensitivity by majority members of the academic community who just don’t “get it” regarding the experiences those students have in academia.[3,7] Collectively, those negative encounters can be perceived by HU students as an assault on their dignity. Social isolation and, frequently, departure from the academic world are the results.[3]

The participation and leadership by Cal-Bridge scholars and alumni in the program’s activities associated with the June 2020 Strike for Black Lives have led to an increase in their sense of belonging and inclusion that we expect will promote their continued academic and professional success.

## Cal-Bridge responds to the Strike for Black Lives

The Cal-Bridge program is a partnership between STEM faculty in the CSU and University of California (UC) systems (see the article by Alexander Rudolph, *Physics Today*, October 2019, page 50). With more than 400 000 undergraduates on 23 campuses, 47% of whom are from URM groups, CSU is the largest, most diverse university system in the nation. The students and alumni of the Cal-Bridge program mirror that diversity, with 59% of the 182 scholars from URM groups and all but one scholar coming from one or more HU groups. The program has had great success. Almost two-thirds of scholars are currently enrolled in PhD programs in the UC system and elsewhere, and most of the rest are enrolled in master's programs with plans to apply to PhD programs eventually. Two of the initial cohort of five scholars in 2014 have received their PhDs; both are now postdocs.

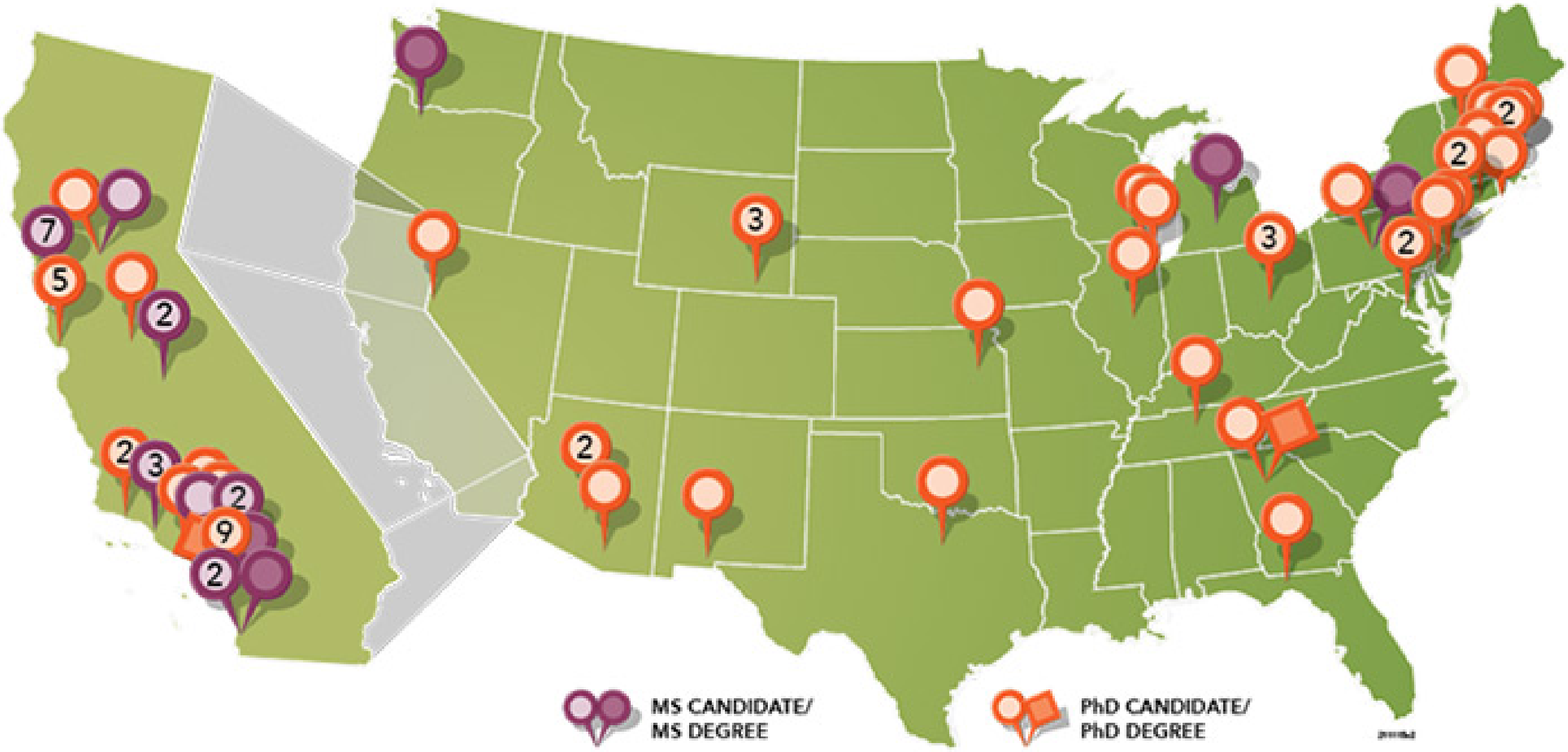

Cal-Bridge scholars have gone on to enroll in graduate schools across the country. Numbers within the pins indicate the total number of scholars in a program when there is more than one. Credit: Cal-Bridge program

In response to the call for a general STEM strike on 10 June 2020, the Cal-Bridge program engaged all levels of the Cal-Bridge community by organizing two hour-long Zoom meetings, one for faculty and staff and another for students and alumni. Twenty-three Cal-Bridge students and alumni (23% of the total at that time) attended the scholars meeting, and 55 faculty and staff attended the faculty meeting. In both meetings the attendees broke into small groups for facilitated discussions of how to respond to the call for antiracist action.

On 24 June a group of more than 50 Cal-Bridge scholars, faculty, program staff, and university administrators gathered for an additional two-hour follow-up meeting to turn the ideas of 10 June into concrete actions. In the first hour, participants broke into groups to discuss the ideas each person was most excited about, and then those groups reported back to the larger group. The larger group identified the top actions to adopt and selected the participants and leaders to pursue each idea. For the second hour, participants broke into groups and turned the ideas into action plans, which would be carried out by the six groups described in the box below.

### Action groups and their missions (* indicates scholar-led)

1. **Outreach to community college and CSU students.*** Develop and coordinate outreach efforts through the Cal-Bridge network and partners, encouraging students, especially Black students, to pursue STEM fields in high schools, community colleges, and universities.
2. **Diverse speaker series.** Invite speakers to showcase future career paths for Cal-Bridge scholars and the contributions of scientists of color (including Cal-Bridge alumni), who can serve as role models. Give students opportunities to interact with those speakers. Discuss anti-racism and experiences of scientists of color. Provide seminar experiences for campuses that don't normally have seminars.
3. **Space for Black mentors and mentees.*** Design space for Black students to acquire skills, knowledge, and confidence through mentoring and practice to be successful in research and coursework.
4. **Mental health.*** Create support groups for scholars and alumni, expert-guided mental health workshops for faculty mentors and scholars, and mental health awareness through compilation of information and resources.
5. **Graduate admissions, inclusion, and hiring.** Coordinate efforts across UC to make graduate admissions and faculty hiring processes more equitable. Increase the number of Cal-Bridge students accepted to UC programs and the number of alumni working as faculty and postdocs at all Cal-Bridge institutions.
6. **HBCU partnerships.** Establish robust collaborations with historically Black college and university programs and faculty to increase HBCU student participation in summer research. Seek funding through a UC-HBCU program as part of a collaborative, multicampus program. Establish processes and resources that support HBCU undergraduate and graduate students in UC and CSU programs.

All the groups created descriptions and planned their initial actions, which can be found on the Cal-Bridge Activism page of the Cal-Bridge program website. More than 40 scholars, program alumni, faculty, and program staff signed up to join one or more of the action groups.

### Taking action

One of the most gratifying outcomes from the initial scholars-only meeting was a clear message that scholars, led by alumni in PhD programs, wanted a role in program leadership beyond the strike activities. There was also a strong hunger for more engagement with the program and with one another. The scholars and alumni took the lead in three of the action groups.

**Action Group 1,** “Outreach to community college and CSU students,” has two strands. One strand was conceived and is led by three alumni from CSU Northridge: Diana Blanco Hernandez (now a PhD student at UC Santa Cruz), Mary Usufzy (a master’s student at CSU Fullerton), and Vidya Venkatesan (a PhD student at UC Irvine). As undergraduates they ran workshops for lower-division CSU Northridge students about the Cal-Bridge program and its summer research program. Although we cannot know if the workshops were the cause, applications from CSU Northridge have increased in the time since those workshops were implemented.

“As cochair of the Cal-Bridge Partnership Committee, I’m actively engaged in creating the communities we want our scholars to be involved in. As a Black woman, seeing my Cal-Bridge community move to support people like me is exactly what these spaces are for. While Cal-Bridge was created to offer academic and career guidance, the absolutely beautiful community that came with it was something I did not expect but am deeply grateful for.”
— *Jordan Ealy, PhD student, University of Maryland*

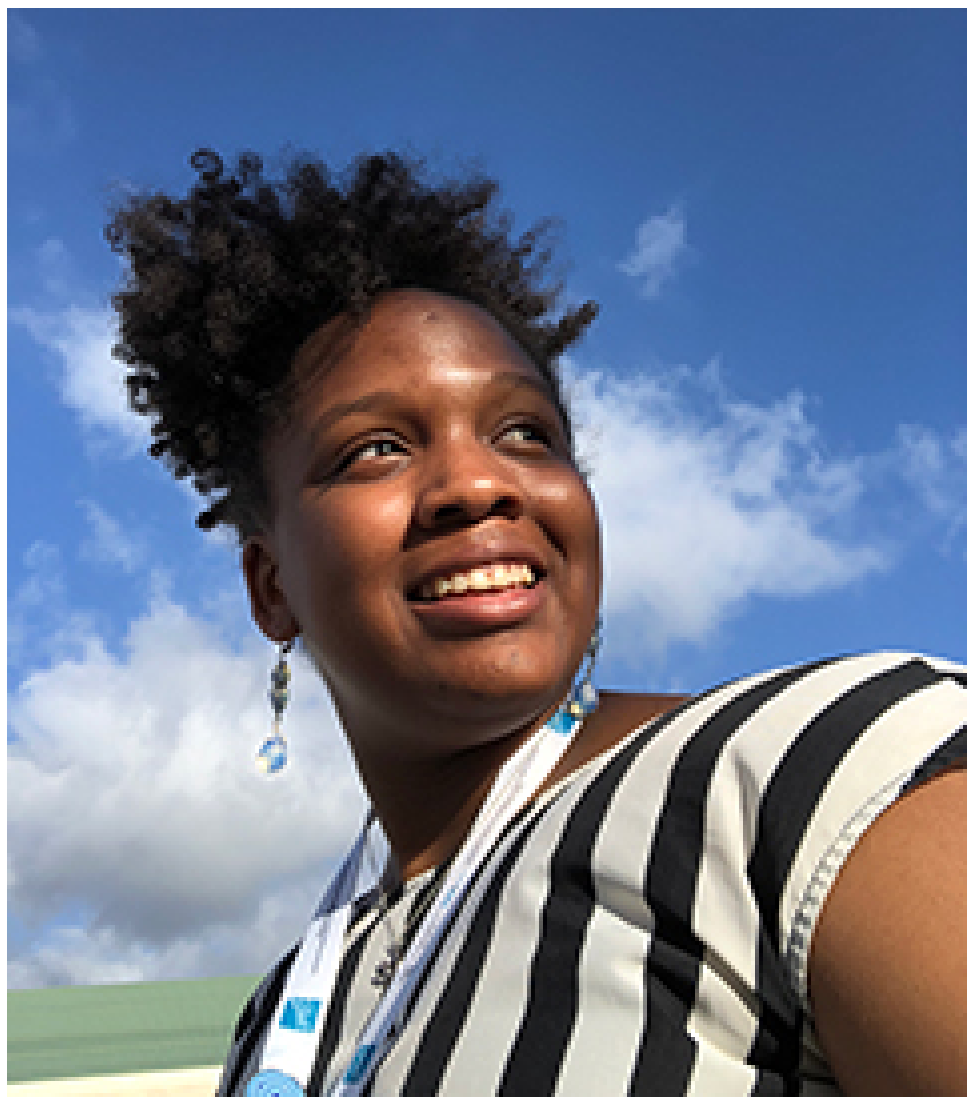

Given their enthusiasm and expertise in recruiting, Blanco Hernandez and Usufzy took on the leadership as cochairs of the Cal-Bridge Outreach and Recruitment committee. Since the 2020–21 academic year the committee has devised and led multiple successful workshops and information sessions on applying to the Cal-Bridge academic year program, the Cal-Bridge Summer program, and other summer research experiences. Although the workshops may not be solely responsible, the number of community college transfer applications has increased since the alumni took over the committee.

The second strand of Action Group 1 resulted from efforts of a second group of alumni, including Roberto Tejada Arevalo (a PhD student at Princeton), Rene Padilla (a PhD student at UC Santa Cruz), and Dennis Calderon-Madera (a PhD student at Ohio State University). As part of the Outreach and Recruitment committee, they are connecting with community college and high school students, particularly Black students, for general recruitment into STEM, not just into the Cal-Bridge program. Many Cal-Bridge scholars come from community colleges and strongly express a desire to help students like themselves follow in their footsteps.

**Action Group 3,** “Space for Black mentors and mentees,” was the brainchild of Don Dixon, a Black Cal-Bridge alum who is now an astrophysics PhD student at Vanderbilt University. The number of Black students in academia is very low, especially in PhD programs, and Dixon wanted to create a space where Black scholars could meet to share their experiences and support one another. Dixon had the help of Shontrice Coleman, a Black staff member of the program who has a BS in physics. Calling themselves the BLack Academic Support and Advisory gRoup (BLASAR), the group’s members have been meeting biweekly since the strike (see the accompanying sidebar by Dixon). Slack and WhatsApp channels facilitate communication between meetings. The group has expanded to include Black STEM PhD scholars from other parts of the country, primarily because of Dixon’s connections to alumni of the Fisk–Vanderbilt Masters-to-PhD bridge program, the leading producer in the country of Black PhDs in multiple STEM fields.[10]

**Action Group 4,** “Mental health,” was the idea of Cal-Bridge scholar Madalyn Johnson (then an undergraduate at CSU East Bay, now a PhD student at UC Santa Cruz). With a small amount of help from a faculty member, Johnson compiled a comprehensive Google spreadsheet that includes mental health resources at each CSU institution, community resources in multiple geographic areas, and a general mental health resource list. In addition, Johnson worked with Katy Rodriguez Wimberly (PhD, UC Irvine; now a postdoc at UC Riverside), director of the Cal-Bridge Peer Mentorship program, to organize peer mentoring workshops on nutrition and physical activity—topics closely tied to mental health —led by a local expert.

### Creating accountability

The organizers of the Strike for Black Lives emphasized the importance of individuals and organizations holding themselves accountable for their proposed actions. A follow-up accountability meeting was held in January 2021, approximately six months after the initial meetings. At this meeting, each of the leaders of the six action groups reported to a gathering of 34 scholars and faculty about their progress in carrying out their proposed activities.

“As a student leader in Cal-Bridge, I brought back what I had learned from these meetings and opened up discussions with others in my department. We continue to talk about current social issues and how we can do our part in acknowledging and addressing them.”
*— Cristilyn Gardner-Watkins, PhD student, University of Wyoming*

Of the six action groups, four reported substantial progress toward the goals they had set. Significantly, three of the four groups were led by students, alumni, or both. The one faculty-led action group reporting significant progress between June 2020 and January 2021 was **Action Group 2**, which sought to create a seminar series of diverse speakers. The committee successfully ran a biweekly series of Zoom talks throughout the 2020–21 and 2021–22 academic years. Most of the talks were recorded and can be viewed on the Cal-Bridge YouTube channel as a resource for HU scholars across the country.

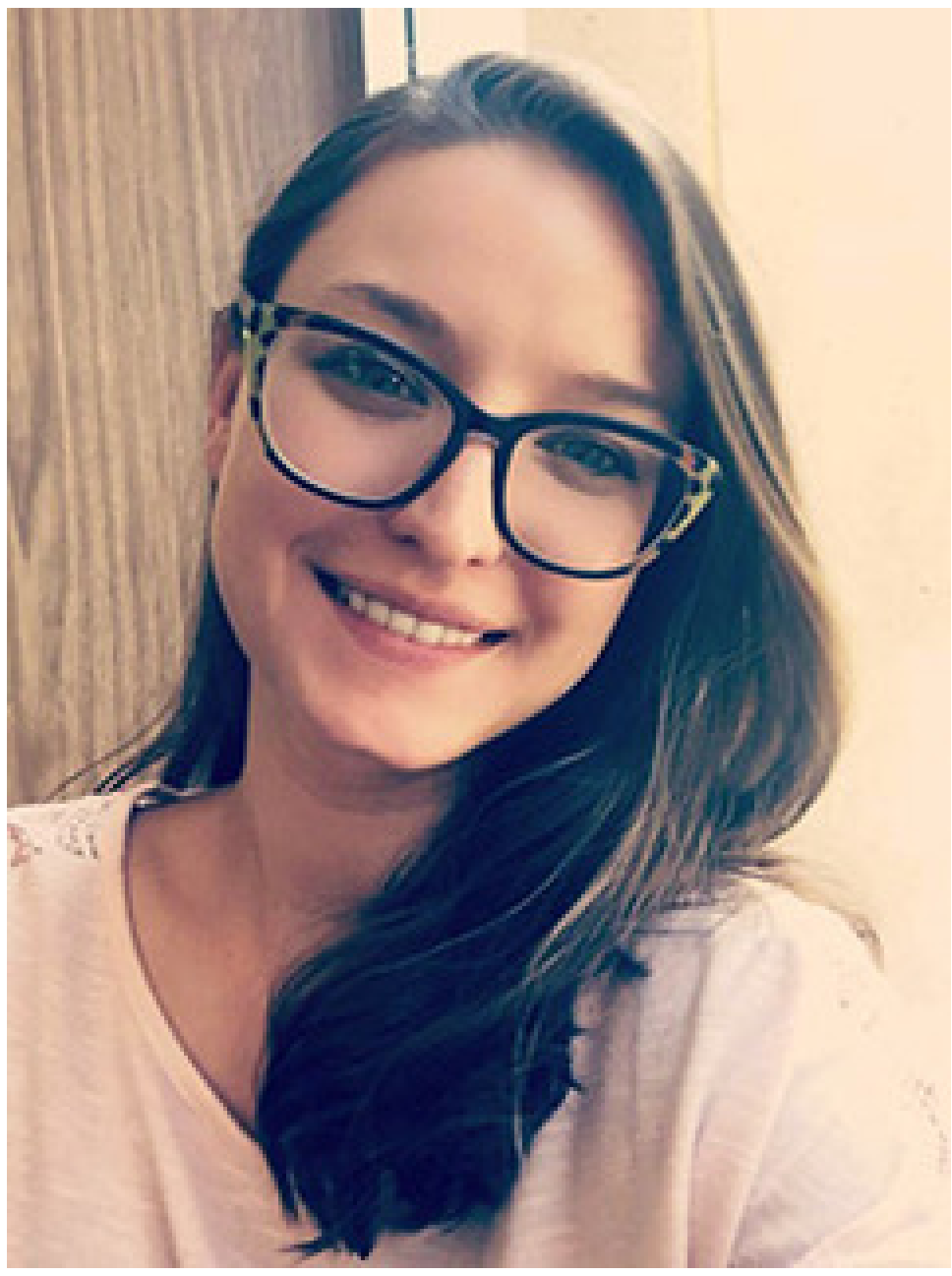

At the January 2021 meeting, after all six action groups had reported, six Zoom breakout rooms, one for each action group, were created, and participants were invited to join one or more of them. The goal was to develop each action group's plans for the next six months.

All six groups came back with clear, concrete plans so that they could be held accountable at future meetings. Significantly, the two action groups that had not been active between June 2020 and January 2021 came back with plans that have subsequently borne fruit. **Action Group 5**, "Graduate admissions, inclusion, and hiring," held a meeting in April 2021 of interested UC PhD admissions directors about best practices for promoting admission of a more diverse population to their PhD programs. Ideas discussed included the use of rubrics and interviews, as well as mechanisms for assessing applicants without the use of standardized GRE (Graduate Record Examination) test scores. The long-term goal is to create an ongoing mechanism for information sharing and group brainstorming about how to improve the representation of HU groups in physics and astronomy PhD programs.

## A forum for Black STEM students

As an aspiring Black astronomer and alumnus of the Cal-Bridge program, I was delighted to see the Cal-Bridge community organize itself to take measurable and meaningful action against anti-Blackness in academia. The emphasis on student involvement made me particularly optimistic that significant and positive changes would occur. In a desire to facilitate those changes, I am leading the BLack Academic Support and Advisory gRoup (BLASAR) to help support Black academics pursuing STEM careers.

The inspiration for BLASAR came during a Cal-Bridge Day of Action event in which several members of the community commented that an organized space for Black students to support one another would be a powerful thing to have. The idea resonated with me, as I attribute many of my struggles during my early academic career to isolation and a lack of guidance. I imagine I would have benefited greatly from such a space—and still would today. As the most senior Black Cal-Bridge alum, I felt taking a leadership role in its creation was an opportunity for me to make a substantial difference in the lives of students like myself.

The group was amorphous in the beginning, but we adapted based on what I saw was helpful to members of the group. Over time I found that our virtual group meetings every other week served as a cathartic release for members when dealing with difficult issues, as a place to share and celebrate their achievements, and as an information hub for career opportunities. For a next step, I decided to extend BLASAR's connections outside the Cal-Bridge network and, by extension, to bolster the networks of individual members.

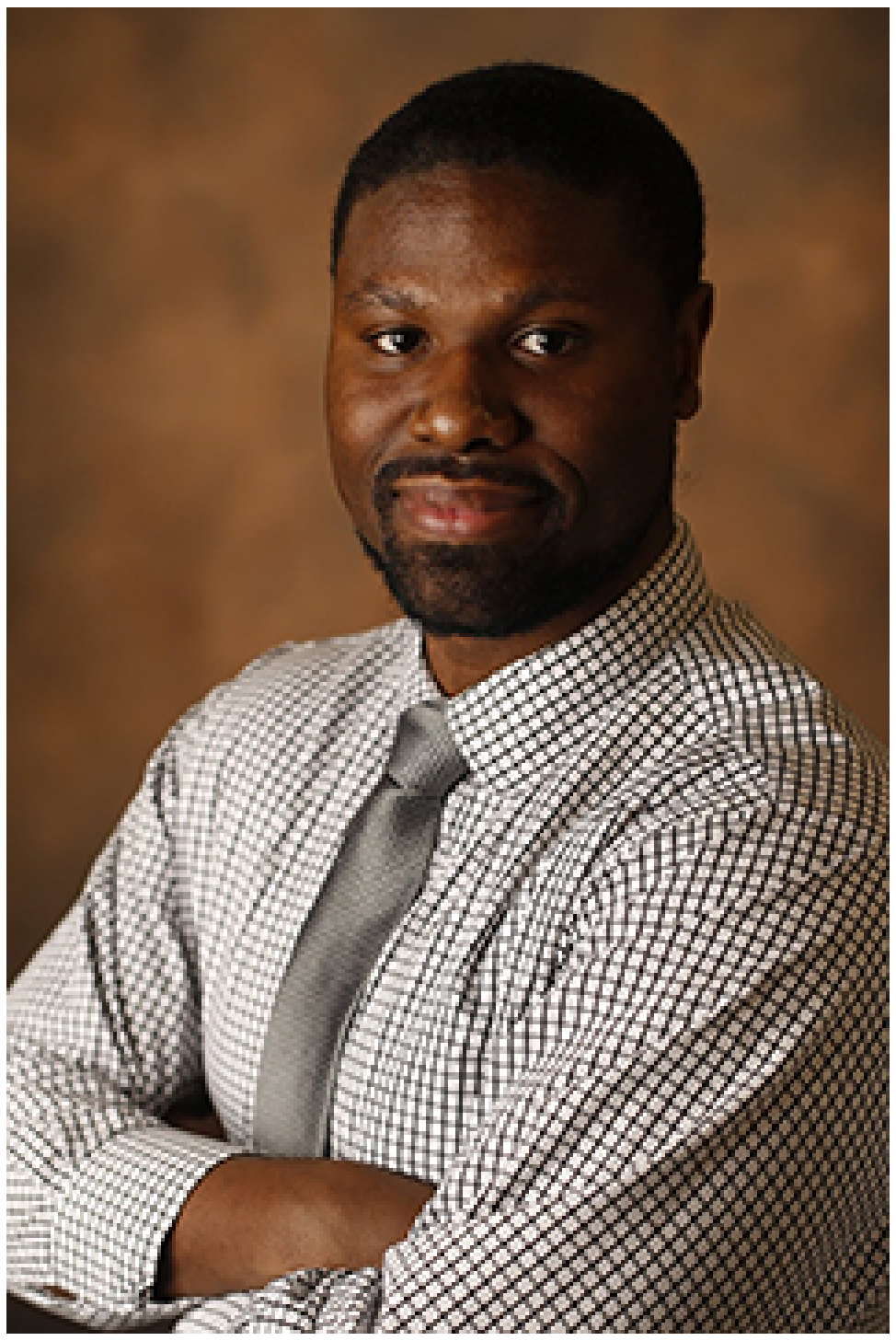
Don Dixon

Now BLASAR is working to form strong ties to other bridge programs and institutions. Among our roughly 15 active members are students from the Fisk–Vanderbilt Bridge Program, American Geophysical Union Bridge Program, NASA, and more. BLASAR is focused on continuing to build strong and mutually beneficial external relationships. We are also actively working to acquire funding to host in-person events and to develop a website. Considering the progress that has been made in the last year, I am looking forward to what will be accomplished in the near future.

— Don Dixon, PhD student, Vanderbilt University

**Action Group 6**, "HBCU partnerships," subsequently organized and submitted a proposal to the UC-HBCU program, administered by the UC Office of the President. A group of UC faculty, joined by the Cal-Bridge director, devised a program whereby students from HBCUs would apply to conduct summer research at one of the 11 physics and astronomy departments in the UC system; those scholars would be designated Cal-Bridge Affiliates and would also participate in the program's academic year activities. Although the proposal was not funded, it provides a blueprint for a potentially significant expansion of the Cal-Bridge program working with HBCU programs and students nationally.

The entire group held additional accountability meetings in June 2021 and June 2022, during which leaders of each action group reported on their progress and breakout discussions were held to determine next steps. Accountability meetings will continue to be held in June of each year.

## Growth of community

Through their participation in the action groups, 21 alumni (37.5% of Cal-Bridge alumni in PhD programs) have become actively involved in the operations of the Cal-Bridge program, including 14 who have taken leadership roles. Rodriguez Wimberly is now cochair of the Mentorship committee, which is responsible for all mentorship training in the program. The alumni leaders of Action Group 1 have become cochairs and members of the Outreach and Recruitment committee. The group also conducts separate outreach to community colleges, in particular California community colleges with large Black enrollments. Other alumni have joined a new Partnership committee, including cochair Jordan Ealy (a PhD student at the University of Maryland). The new committee is defining an application and vetting procedure for non-UC PhD programs to become Cal-Bridge Partners.

"The Strike for Black Lives events led to a more careful examination in our department of diversity among our undergraduate majors. At our majority Hispanic/Latinx campus, we found the percentage of Hispanic/Latinx students in the physics major was very similar to the overall undergraduate population, whereas the number of Black students was somewhat smaller than the undergraduate population in general. That has helped to focus our efforts about recruitment and climate in the major."
*— David Strubbe, Assistant Professor of Physics, UC Merced*

Significantly, alumni who have taken leadership roles in the program report that their sense of community has increased substantially due to their activities. Their feedback indicates that the opportunity to lead these efforts has also had a meaningful impact on their sense of social inclusion and sense of belonging, which will promote their continued academic and professional success.

Faculty who led and participated in Strike for Black Lives Action Groups also reported that their participation profoundly affected their own view of anti-Black racism in their departments and fueled positive action, including departmental self-assessment and new initiatives. Sadly, one faculty member was reported to their administration by an anonymous colleague for participating in the Strike for Black Lives action groups. Ultimately, their experience catalyzed them to positive action within and outside their department. The unfortunate reaction of some colleagues underscores the challenges in creating change.

We hope that this account of the Cal-Bridge community answering the call of the Strike for Black Lives demonstrates the enormous potential for combating anti-Black racism within the physics community and inspires similar actions across STEM. If initiatives such as those reported here are undertaken broadly, we can hope to see a move toward a more just, equitable, and inclusive science community.

*Alexander L. Rudolph is a professor of physics and astronomy at California State Polytechnic University in Pomona and the director of the Cal-Bridge program. Carol Hood is a professor of physics at California State University, San Bernardino, and the associate director of the Cal-Bridge program.*

*The authors would like to thank the members of the Cal-Bridge community who stepped up to participate in the Cal-Bridge Strike for Black Lives activities described in this article. We acknowledge that without those participants, scholars, alumni, faculty, and staff, none of the work described here would have happened. We especially want to thank and acknowledge the Cal-Bridge scholars and alumni who took on leadership positions within the Cal-Bridge community and helped promote the increased sense of cohesion and community reported in this article: Katy Rodriguez Wimberly, Eleanor Lyke, Don Dixon, Cristilyn Gardner-Watkins, Justin Bracks, Evan Haze Nuñez, Dennis Calderon-Madera, Diana Blanco Hernandez, Jordan Ealy, Rene Padilla, Roberto Tejada Arevalo, Mary Usufzy, Vidya Venkatesan, and Madalyn Johnson.*